\documentclass[onecolumn]{aa}  

\usepackage{xcolor}
\usepackage{float}
\usepackage{comment}
\usepackage{graphicx}  
\usepackage{gensymb}
\usepackage[varg]{txfonts}
\usepackage{hyperref}
\hypersetup{
    colorlinks=true,
    linkcolor=blue,
    filecolor=magenta,      
    urlcolor=cyan,
   citecolor=blue,
}
\newcommand{\dd}{\ensuremath{\mathrm{d}}}

\newcommand{\CaII}{\ion{Ca}{II}}
\newcommand{\CaIIH}{\CaII\ H}
\newcommand{\CaIIHK}{\CaII\ H \& K}
\newcommand{\CaIIK}{\CaII\ K}
\newcommand{\CaIIIR}{\CaII\ 854.2~nm}
\newcommand{\Icont}{\ensuremath{I_\mathrm{cont}}}

\begin{document}

   \title{The magnetic sensitivity of the Ca II H and K lines}

   \author{M. Kriginsky\inst{1}, J. Leenaarts\inst{1}, J. de la Cruz Rodr\'\i guez\inst{1}, S. Danilovic\inst{1}, O. Andriienko\inst{1}}

   \institute{Institute for Solar Physics, Dept. of Astronomy, Stockholm University, AlbaNova University Centre, 106 91, Stockholm, Sweden
         }

   \date{Received ; accepted }

 
  \abstract
{The solar chromosphere is a transition layer between the cool, dense photosphere and the hot, rarefied corona. This boundary region plays a key role in regulating energy transport and structuring the magnetic field throughout the solar atmosphere. Understanding its thermodynamic and magnetic properties is essential to model and interpret solar phenomena.}
{This study investigates the theoretical properties and diagnostic potential of the polarisation signals in the \CaIIHK\ lines, with particular emphasis on their capability to probe magnetic fields in the upper chromosphere with the CHROMIS instrument at the Swedish 1-m solar telescope.}
{We combine semi-empirical atmospheric models with high-resolution solar observations to model the formation of the \CaIIHK\ lines using non-local thermodynamic equilibrium
 radiative transfer calculations. The sensitivity of the lines to the magnetic field is examined through response functions and synthetic inversions, enabling an assessment of their diagnostic performance under realistic chromospheric conditions.}
{For typical chromospheric field strengths, the linear polarisation of the \CaIIHK\ lines is less than 1.7\%, below the expected detection threshold of CHROMIS. However, their circular polarisation reaches more than 10\% in strong-field regions, which is detectable by CHROMIS. Both lines are sensitive to magnetic fields in the upper chromosphere, with the K line forming slightly higher due to its larger opacity, and the H line exhibiting a somewhat stronger Zeeman sensitivity owing to its higher effective Landé factor and longer wavelength. Using the weak-field approximation, the line of sight magnetic field can be reliably inferred around $\log{\xi}\!\approx\!-4$. These results confirm that the \CaIIHK\ lines constitute powerful diagnostics for studying the magnetic structure of the upper solar chromosphere.}
{}


   \titlerunning{}
   \authorrunning{M. Kriginsky et al.}
   \maketitle
   \nolinenumbers
   
%

\section{Introduction}

Among the many challenges in studies of the solar chromosphere, inferring the magnetic field is one of the most critical and elusive tasks. Magnetic fields play a pivotal role in shaping the energy distribution and dynamic behaviour of the solar atmosphere. Unfortunately, many chromospheric lines suffer from small effective Landé factors, which reduce their sensitivity to the Zeeman effect and complicate magnetic diagnostics. Additionally, few magnetically sensitive lines form in the chromosphere, with the most commonly used line being the central line of the infrared triplet of \ion{Ca}{II} at 854.2 nm, which has been the workhorse for chromospheric studies, both in the past \citep[][]{1991A&A...247..379W,2007ApJ...663.1386P,2012ApJ...748..138K,2012SoPh..280...69H} and in more recent studies, using high-resolution observations \citep[][]{2020A&A...642A..61K,2021A&A...650A..71K,2020A&A...642A.210M,2021A&A...649A.106Y,2022A&A...662A..88V}. This line has a moderate Land{\'e} factor of 1.1, and its formation can be modelled without including complex phenomena such as partial frequency redistribution (PRD) of photons. Additionally, its location in the infrared spectrum provides an advantage as a result of the increased Zeeman splitting at longer wavelengths, enhancing sensitivity to magnetic fields.

The \ion{Ca}{II} 854.2 nm line offers several practical benefits for diagnostics. It forms over a relatively broad range of heights, spanning the upper photosphere to the lower chromosphere, and the relatively expanded magnetic field at these heights allows the application of simplified diagnostic techniques such as the weak-field approximation \citep[WFA,][]{2004ASSL..307.....L}.

However, despite its utility, the \ion{Ca}{II} 854.2 nm line has limitations. Its sensitivity to magnetic fields peaks in the upper photosphere and lower chromosphere \citep[][]{2016MNRAS.459.3363Q} makes it less effective for probing magnetic fields in the middle and upper chromosphere. Therefore, the use of additional spectral lines that form at higher layers is desirable. These lines, while potentially more challenging to model and interpret, can provide the sensitivity needed to diagnose magnetic fields in the upper reaches of the chromosphere.

The \CaIIHK\ lines at 396.85 nm and 393.37 nm are potentially suitable lines, as they are formed in the middle and upper chromosphere. They are probes of the thermodynamic and kinetic properties of the middle and upper chromosphere \citep{2018A&A...611A..62B}. An alternative set of diagnostics, which has received considerable attention, is the \ion{Mg}{II} h and k lines around 280 nm. The \ion{Mg}{II} lines have served as the target for magnetic field diagnostics in the chromosphere \citep[][]{2021SciA....7.8406I,2022ApJ...933..145L,2024ApJ...975..110L}. However, they can only be observed from outside the Earth's atmosphere, requiring satellite missions or rocket experiments. In contrast, the \ion{Ca}{II} H and K lines can be observed from the ground, making them more suitable for routine spectropolarimetric campaigns.

However, the use of \ion{Ca}{II} H and K lines in spectropolarimetry remains under-explored 
\citep{1990ApJ...361L..81M}.
This limitation is due to a combination of factors. First, the photon flux in the blue part of the spectrum is lower compared to the red part, reducing the S/N in observations. Second, the effective Land{\'e} factors of these lines 
($1.169$ for the \ion{Ca}{II} K line and $1.33$ for the \ion{Ca}{II} H line ) suggest that their sensitivity to magnetic fields may be modest. However, in active regions, the line profiles vary strongly, which corresponds to steep intensity gradients \citep[$\dd I / \dd \lambda$, e.g.][]{2018A&A...612A..28L}. As a result, the circular polarisation signal can still reach more than 10\% of the intensity, despite the modest Land{\'e} factor 
\citep{1990ApJ...361L..81M}.

The polarisation of the \CaIIHK\ lines is subject to the effects of phenomena that are neglected in most inversion studies, such as atomic polarisation and J-state interference \citep{2011ApJ...743....3B,2025A&A...704A.173J}. Some properties of the linear polarisation seen in these lines can only be explained by accounting for these phenomena. An example is the zero-level crossing of the linear polarisation fraction between the core of both lines described by
  \citet{1980A&A....84...68S}, 
  which can be reproduced when quantum interferences are accounted for 
  \citep{2011ApJ...743....3B}.

Recent and forthcoming advances in instrumentation are likely to renew interest in the spectropolarimetric signal of the \CaII H and K lines. Spectropolarimeters capable of observing the blue part of the visible spectrum have already been developed. For example, the Sunrise UV Spectropolarimeter and Imager \citep[SUSI,][]{2020SPIE11447E..AKF} instrument on board the SUNRISE III balloon experiment \citep[][]{2025SoPh..300...75K} and the Visible Spectro-Polarimeter \citep[ViSP,][]{2022SoPh..297...22D} at the Daniel K. Inouye Solar Telescope \citep[DKIST,][]{2020SoPh..295..172R} are tailored for such studies. The CHROMospheric Imaging Spectrometer \citep[CHROMIS;][]{2017psio.confE..85S} at the Swedish 1-meter solar telescope \citep[SST,][]{sst} was upgraded in the spring of 2025 to allow for full-Stokes spectropolarimetry in \CaIIHK. It is undergoing calibration and testing as of the summer of 2025. 

The purpose of this study is to investigate the magnetic sensitivity of the \ion{Ca}{II} H and K lines, along with their potential use for magnetic field diagnostics. This goal is primarily motivated by the advantages and disadvantages of CHROMIS. It is a dual Fabry-P{\'e}rot-based imaging spectropolarimeter for the 390-500 nm range, and can operate simultaneously with the CRISP2 
instrument
\citep{2026A&A...705A..55S}.
The latter is a new instrument for the 500-900 nm range that replaced the old CRISP instrument
\citep{2008ApJ...689L..69S}
in the summer of 2025. Both instruments can obtain images with a diffraction-limited spatial resolution. One scientific goal of these instruments is to observe multiple spectral lines together and use these observations as input to multi-line multi-spatial-resolution inversions
\citep{2019A&A...631A.153D}
in order to produce full 3D models of the thermodynamic and magnetic properties of the solar photosphere and chromosphere. Adding spectropolarimetric capabilities to CHROMIS helps to constrain the magnetic field in the middle and upper chromosphere.

Because of the low photon flux in the \CaIIHK\ lines and the need to scan wavelengths in the line consecutively, we do not expect to achieve an S/N larger than $\sim200$ without sacrificing too much in temporal cadence. We thus want to investigate whether we can expect to observe this
linear polarisation at all because linear polarisation signals tend to be weaker than circular polarisation.
Given this limitation, we also want to investigate whether it is required to include 
atomic polarisation and J-state interference. Adding those processes adds orders of magnitude to the time required for inversions
\citep[on the  order of $10^3$ hour per pixel,][]{2022ApJ...933..145L}
compared to inversions that only include the Zeeman effect, which take less than one hour per pixel. 

Finally, given the large field of view (FOV) of CHROMIS ($80\times80$ arcsec corresponding to $\sim 2200\times2200$ pixels), even a Zeeman-effect-only inversion of a single line scan would take on the order of $10^6$ CPU hours. Therefore, we also investigate the accuracy of the WFA in these lines, as this would give a quick way to get a rough idea of the magnetic field that can be run on a small computer.

To do this, we start with investigating the expected polarisation signal in a semi-empirical model atmosphere for atomic polarisation, J-state interference, and the Zeeman effect. We will show that for the expected S/N of CHROMIS observations, we can only expect to see circular polarisation signals caused by the Zeeman-effect. Then we create a 3D empirical model atmosphere from which we compute synthetic \CaIIHK\ spectra. We invert these spectra with the weak-field approximation to investigate the accuracy of the derived magnetic field compared to the ground truth.

\begin{figure}
 \centering
 \includegraphics[width=9cm]{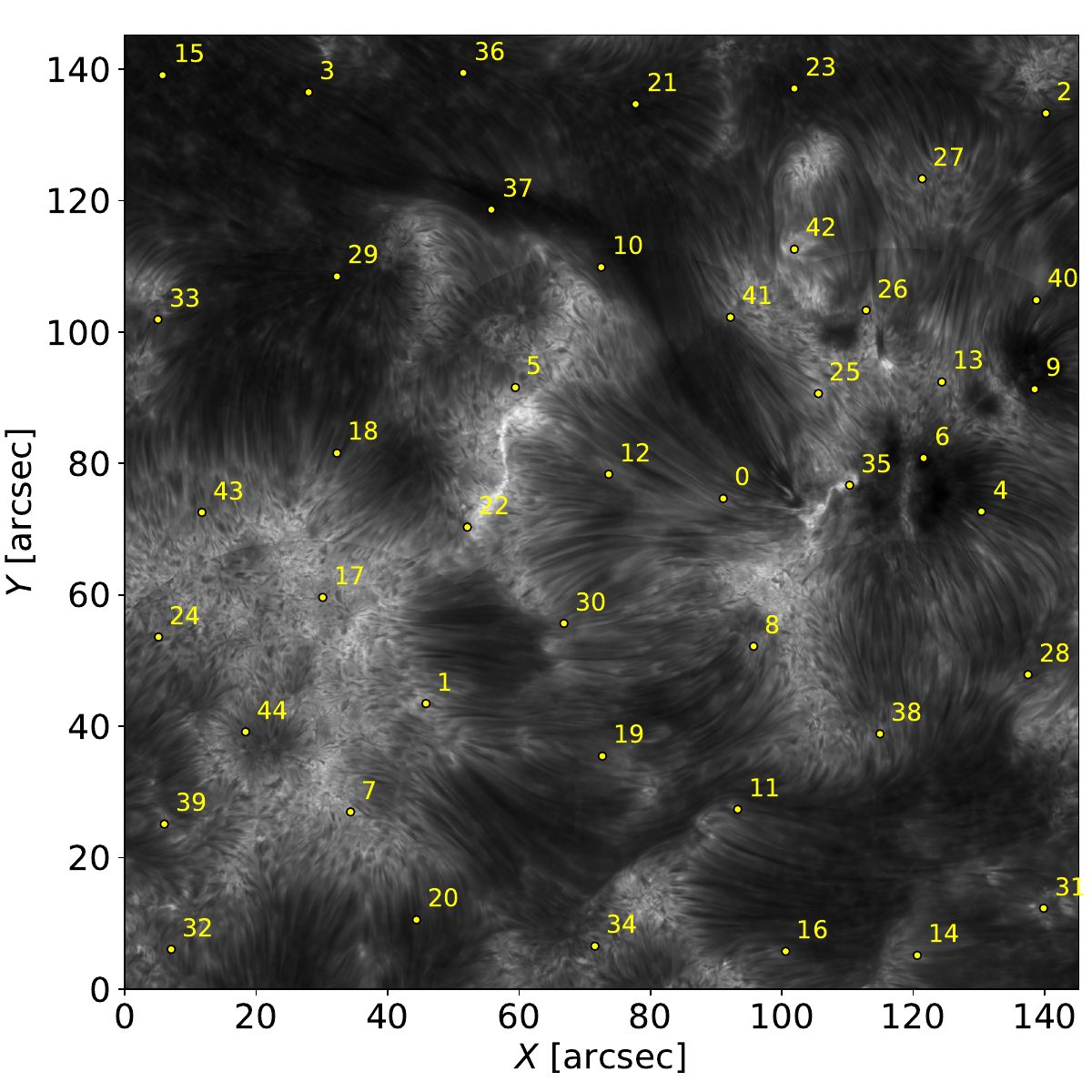}
 \caption{Image of active region NOAA 13465 in the line core of the \CaIIIR\ line. The marked pixels are the locations of 45 representative line profiles that sample the variety of line profiles present in the observations as determined by $k$-means clustering (see Sec.~\ref{subsec:inversion_strategy}).}

 \label{Figure:1}%
\end{figure}
 
 \section{Methods} \label{sec:methods}

\subsection{Synthesis code} \label{synth}

One of our goals is to determine the importance of applying the correct theoretical formalism when solving the line formation problem for the \CaIIHK\ lines. For this purpose, we employ the HanleRT code \citep[][]{2016ApJ...830L..24D}, which can solve the line formation problem accounting for the joint action of the Zeeman effect,  J-state interference, and atomic polarisation mechanisms for arbitrary magnetic fields. HanleRT incorporates the multi-term formalism of \citet{2014ApJ...791...94C}, including the effects of possible coherences that are neglected when using a multilevel formalism, while also accounting for the effects of PRD. We use this code to generate spectral profiles under the multilevel and multiterm approximations, to determine when the multilevel approach is sufficiently accurate in reproducing the polarimetric properties of these lines.

\subsection{Observations} \label{subsec:observations}

For our study, we would ideally use real spectropolarimetric observations in the \CaIIHK\ lines obtained with instruments such as CHROMIS. However, these data are not available yet and we needed to consider an alternative. We decided to use atmospheric models obtained from spectropolarimetric inversions applied to observations. We used this approach instead of performing the synthesis of data from numerical simulations because simulation spectra have been shown to be considerably different from observed ones
\citep{2018A&A...611A..62B}.

We used observations obtained on 17 October 2023 with the CRISP imaging spectropolarimeter \citep[CRISP]{Scharmer_2008} at the SST starting at 10:26:10 UT. The observations were performed in a 3x3 mosaic mode, with an overlap of 30\%, providing an overview of the active region NOAA 13465 as it neared the disc centre. Spectropolarimetric observations of the \ion{Ca}{II} 854.2 nm line were obtained at 15 wavelength positions, sampling the interval [$-6.75$, $6.75$] $\times 10^{-2}$ nm around the line core.
Raw data were reduced with the SSTRED
processing pipeline \citep{2015A&A...573A..40D,2021A&A...653A..68L}, including image restoration using the Multi-Object Multi-Frame Blind Deconvolution \citep[MOMFBD;][]{2002SPIE.4792..146L,VanNoort2005} method. 

This particular observation is an ideal target for the current study, as it spans a large FOV near disc centre, including pixels located in the active region sunspot, the surrounding plage areas, and large quiet-Sun regions (see Fig.~\ref{Figure:1}).

\subsection{Inversion code}

We perform spectropolarimetric inversions on the observations of the \CaIIIR\ line to obtain an empirical 3D atmospheric model. We also used inversions of synthetic spectra of the \CaIIHK\ lines computed from this model to determine how accurate the magnetic fields inverted from their spectra can be.

To perform the inversions we used the STockholm inversion Code \citep[STiC,][]{2016ApJ...830L..30D,2019A&A...623A..74D}, an inversion tool that incorporates a modified version of the Rybicki and Hummer code \citep[RH,][]{2001ApJ...557..389U} as its forward solver under the assumption of a 1D plane-parallel atmosphere. Within STiC, the radiative transfer equation is solved using cubic B{\'e}zier solvers \citep[][]{2013ApJ...764...33D}. The inversion process also relies on an equation of state extracted from the Spectroscopy Made Easy code
\citep[SME, ][]{2017A&A...597A..16P},
and includes the effects of PRD following
\citet{2012A&A...543A.109L}.

\subsection{The inversion strategy} \label{subsec:inversion_strategy}

We applied the $k$-means clustering method to the observations described in Sec~\ref{subsec:observations}, obtaining 45 clusters. We then used STiC to invert one profile from each cluster (identified in Fig.~\ref{Figure:1}) to obtain atmospheric models that together form a representative sample of the variety of atmospheres that one might encounter when observing the chromosphere.

For each chosen profile, the FAL-C atmosphere was used as an initial guess, with an ad-hoc vertical velocity stratification added. Each pixel was inverted, assuming either a constant velocity or one varying with depth. The inversions were carried out at 50 equidistant depth points within the column mass range $\log{\xi} \in [-6.5, 1.0]$. The column mass, $\xi$, represents the integrated mass per unit area above a given height, defined as

\begin{equation}
\xi = \int \rho ds,
\end{equation}

where $\rho$ is the mass density, $ds$ is the differential path length along the vertical direction. In centimetre-gram-seconds (cgs) units, $\xi$ is expressed in g cm$^{-2}$.

Multiple inversion cycles were performed, increasing the number of nodes after each cycle. The inverted physical parameters were the temperature $T$, the vertical velocity $v$, the microturbulent velocity $v_{\mathrm{turb}}$, the parallel and perpendicular magnetic field components ($B_{\mathrm{LoS}}$ and $B_{\mathrm{PoS}}$, respectively) and the azimuth angle $\chi$. The number of nodes for each parameter and cycle is given in Table~\ref{table:Table1}.

The first two cycles were devoted to the determination of the stratification of temperature and vertical velocity, while the magnetic field was kept undetermined. The third and fourth cycles were used only for the magnetic inversion, using as an initial guess the value obtained through the use of the weak-field approximation.

For each pixel, we chose the atmospheric model that yielded the smallest value of the $\chi^2$ function, defined as
\begin{equation}
\chi^2 = \frac{1}{4s} \sum_{k=0}^{3} \sum_{i=1}^{s} \left( \frac{I_k^{\text{obs}} - I_k^{\text{mod}}}{w_{k,i}} \right)^2,
\end{equation}
where $w_{k,i}$ is the weight associated with the noise level of the Stokes parameter $k$ at the wavelength index $i$, $s$ is the number of wavelength points, $I^{\text{obs}}$ is the observed Stokes vector, and $I^{\text{mod}}$ is the Stokes vector produced by the forward engine of the inversion code given the input stratification of the inverted physical parameters. 

We then inverted the full FOV consisting of 10,890,000 pixels. To make the task computationally feasible, we adopted the following scheme:

\begin{enumerate}
\item The pixels in the FOV were randomly divided into 40 groups.

\item For the first group, each pixel was initialised using the atmospheric model corresponding to the $k$-means cluster to which it belonged.

\item These pixels were inverted with STiC using three inversion cycles, employing node configurations equivalent to cycles 2, 4, and 5 of Table~\ref{table:Table1}.

\item A neural network was then trained to reproduce the inverted atmospheres from this first group, using the observed Stokes profiles as input. The network architecture was adapted from \citet[][]{2025ApJ...981..121K}, with additional parallel branches to estimate the magnetic field stratification.

\item For each of the remaining pixel groups:
\begin{enumerate}
    \item the neural network (continually updated after each group) was used to predict an initial atmospheric model for every pixel in the group;
    \item these pixels were then inverted with STiC using the same three-cycle scheme as in Step 3;
    \item the neural network was retrained by including the newly inverted pixels, progressively improving its predictions.
\end{enumerate}

\item This process was repeated until all 40 groups had been inverted.
\end{enumerate}

This hybrid inversion–prediction approach, with the neural network acting as an intermediate step, reduced the total inversion time by a factor of at least three.

 \begin{table}
         \renewcommand*{\arraystretch}{1.5}
        \centering
        \caption{ Node distribution for each of the inversion cycles. }
        \label{table:Table1}
        \begin{tabular}{cccccc}
        
            \hline \hline
            
            Parameter&Cycle 1&Cycle 2&Cycle 3&Cycle 4&Cycle 5\\
            \hline
            $T$&5&6&0&0&6\\

            $v$&2&3&0&0&3\\

            $v_{\mathrm{turb}}$&0&2&0&0&2\\

            $B_{\mathrm{LoS}}$&0&0&1&2&4\\

            $B_{\mathrm{PoS}}$&0&0&1&2&2\\

            $\chi$&0&0&1&1&1\\
        \hline \hline
        \end{tabular}
        \tablefoot{The number zero means the value of the parameter is kept constant at its input value.}
\end{table}        

\section{Results}\label{sect:results}

\subsection{Semi-empirical atmosphere} \label{sect:falc}

We compared the multilevel Zeeman approximation (MZ) with the multiterm formalism (MT)
to assess the errors introduced by the simplified MZ approach in reproducing the spectra of the \CaIIHK\ lines. The calculations were carried out using the semi-empirical FAL-C model \citep[][]{1985cdm..proc...67A,1993ApJ...406..319F}, with a constant ad hoc magnetic field. For the MZ approach, we employed a five-level plus continuum \ion{Ca}{II} atom, whereas for the MT approach a three-term \ion{Ca}{II} atom was considered (Del Pino Aleman, private communication). In both cases, PRD was included in the \ion{Ca}{II} H and K lines, while the infrared triplet lines were treated under the assumption of complete redistribution (CRD).

\subsubsection{Non-magnetic case}

\begin{figure}
 \centering
 \includegraphics[width=8cm]{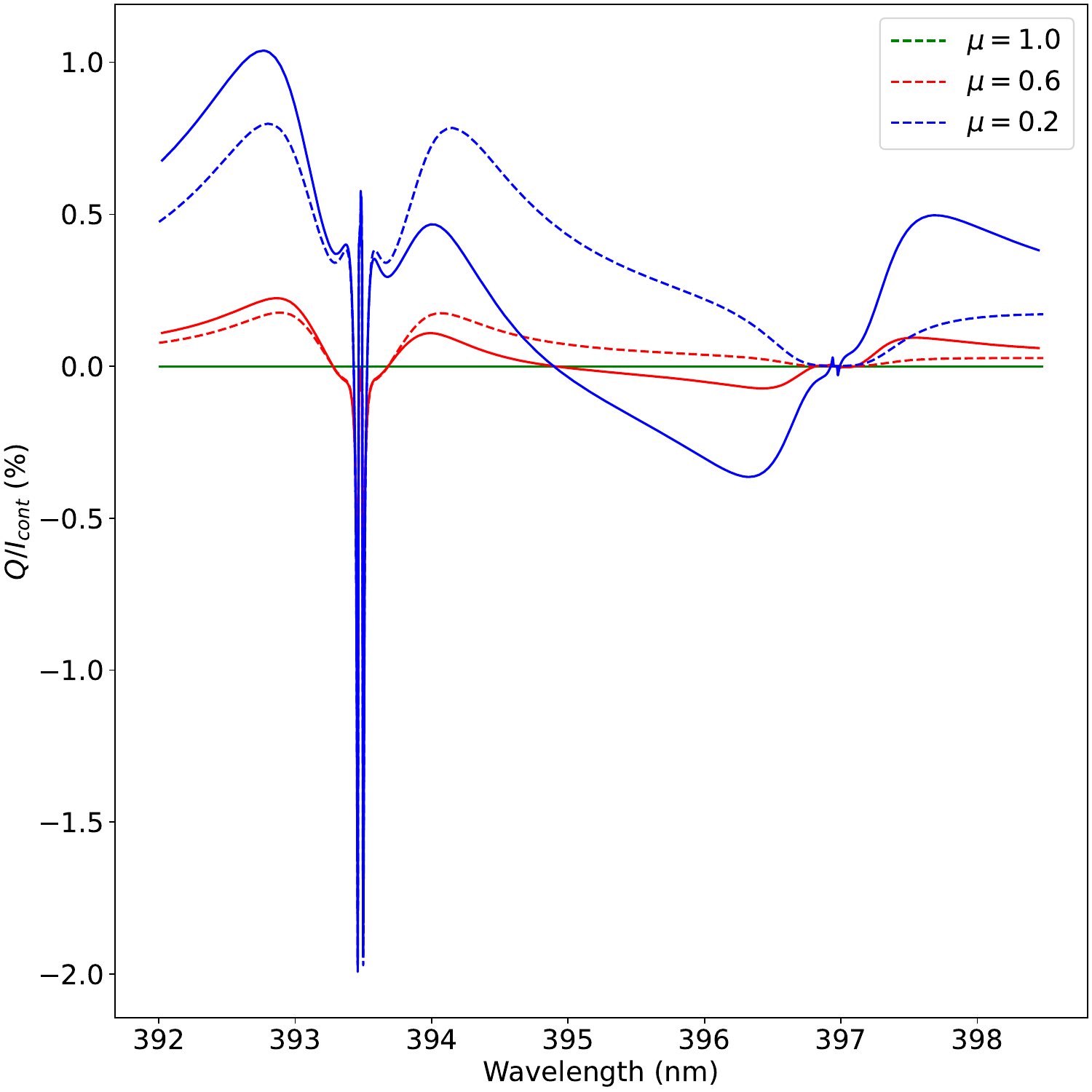}
 \caption{Comparison of linear polarisation of the \ion{Ca}{II} H and K lines computed with the multiterm (MT, solid curves) and multilevel (ML, dashed curves) formalisms for different values of the heliocentric angle $\mu$ for the case of zero magnetic field. The colours represent the different values of $\mu$ as given in the legend. For $\mu=1$ no scattering polarisation is produced and both curves are identically zero. }
 \label{Figure:2}%
\end{figure}

We first considered the formation of fractional linear polarisation profiles in the absence of magnetic fields. For this test, we fixed the azimuth angle so that $U = 0$. Figure~\ref{Figure:2} compares the results of the MZ and MT approaches for different values of the heliocentric angle $\mu$. The MT approach  reproduces the expected zero-level crossing between the cores of the two lines and the small lobes around the core of the \ion{Ca}{II} H line, already reproduced in previous studies \citep[][]{2011ApJ...743....3B,2025A&A...704A.173J} for $\mu<1$. Because the atmosphere is assumed to be plane-parallel, no linear polarisation is produced for $\mu=1$. As expected, the MZ approach cannot reproduce the effects of quantum interferences, as those are neglected, and predicts incorrect linear polarisation in the wings of the lines. However, both approaches yield nearly identical results for the core of the \ion{Ca}{II} K line.

\subsubsection{Magnetic case}

\begin{figure*}
 \centering
 \includegraphics[width=18cm]{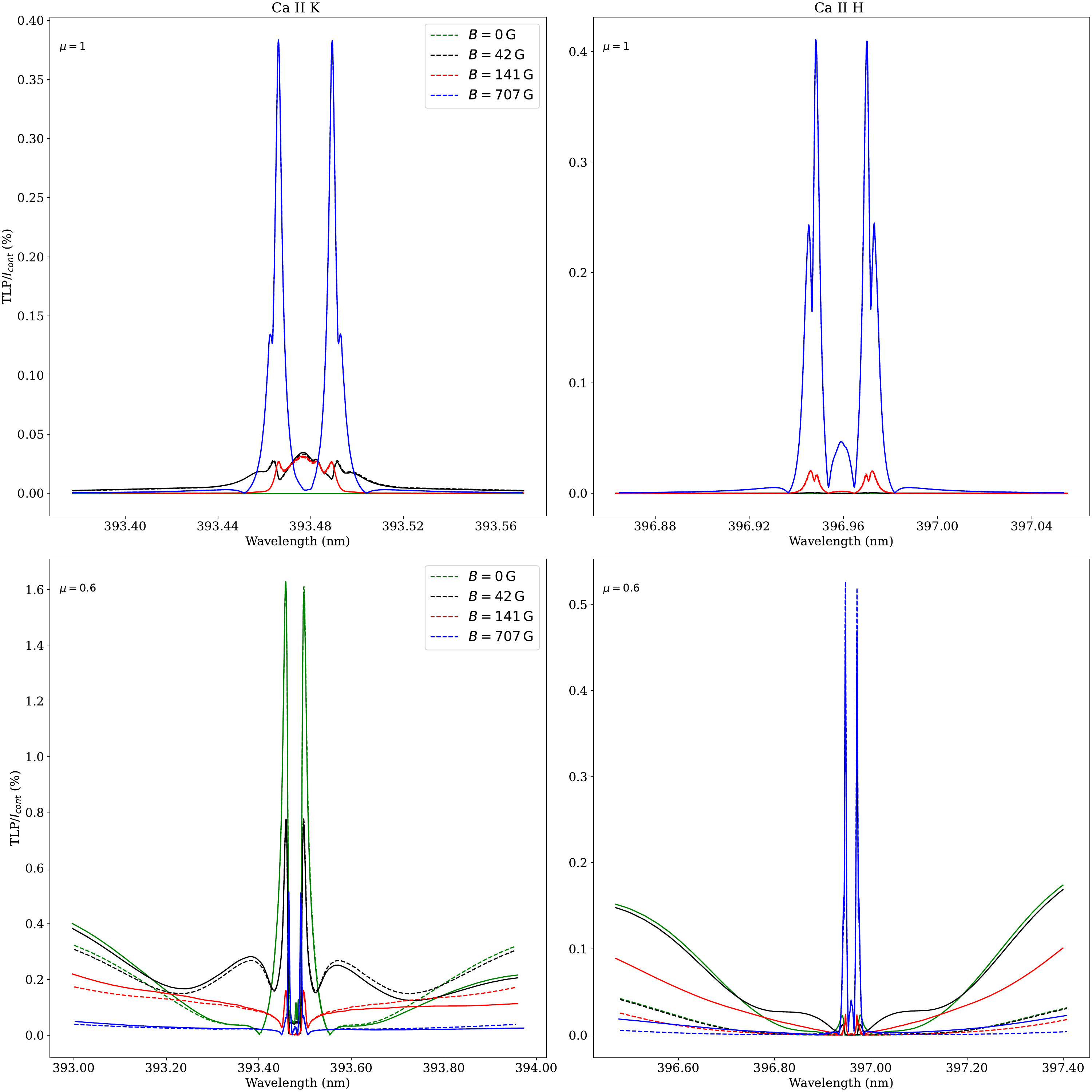}
 \caption{Fractional linear polarisation of the \ion{Ca}{II} H and K lines under the multilevel and multiterm formalisms. As in Fig.~\ref{Figure:2}, continuous curves represent the results of the MT formalism, while the results of the MZ formalism are shown in dashed curves. The different colours represent different values of $B$, which is shown inside the panels. The top panels show the results viewed at disc centre ($\mu=1$), while the bottom panels show the results for $\mu=0.6$. The panels on the left show the results for the \ion{Ca}{II} K line, while the panels on the right side show the results for the \ion{Ca}{II} H line.}
 \label{Figure:3}
\end{figure*}

To evaluate the effects of the presence of a uniform, non-zero magnetic field on the fractional polarisation under both the MZ and MT formalisms, we performed a spectral synthesis for different values of the total magnetic field strength ($B$). The magnetic field inclination with respect to the vertical direction was kept fixed at 45\degree. Fig.~\ref{Figure:3} shows the fractional total linear polarisation
\Big($\mathrm{TLP}/\Icont = \sqrt{Q^2 + U^2}/\Icont$\Big)
under both formalisms for different values of the total magnetic field strength and heliocentric angle.

We start by looking at the result for $\mu=1$. Because the model atmosphere is plane-parallel, there is no scattering polarisation, and the non-zero TLP is caused by the Zeeman effect. The signals are weak, and even at $B=707$~G (i.e the horizontal component of the field is 500~G), the maximum $\mathrm{TLP}/\Icont$ is only 0.4\%.

Towards the limb (bottom panels), the magnetic field has a depolarising effect, with the value of $\mathrm{TLP}/\Icont$ decreasing in both the line core and the line wings with an increasing magnetic field. The maximum $\mathrm{TLP}/\Icont$ (for zero field) is above 1.5\%, decreasing to 0.8\% for a field of 42~G, and we speculate that TLP signals originating from weak-field regions might just be visible with CHROMIS.

The agreement between both formalisms is good around the core of the \ion{Ca}{II} K line. The fact that the \ion{Ca}{II} H line is intrinsically unpolarisable means that the MZ modelling will be unable to reproduce the correct line profiles for magnetic fields that do not give rise to significant Zeeman-induced linear polarisation.

Based on the results of Fig.~\ref{Figure:3} we do not expect to see any TLP (i.e. Stokes $Q$ and $U$) at disc center outside of regions with very strong horizontal fields (such as penumbras). That is because for an instrument like CHROMIS, whose maximum expected S/N level is 200 for the intensity, fractional linear polarisation levels below 0.5 \% are expected to be below the detection threshold. Therefore, if the observational target is a weak-field region, observed Stokes $Q$ and $U$ will likely be an effect of calibration errors in the observations.

\subsubsection{Circular polarisation}

\begin{figure*}
 \centering
 \includegraphics[width=18cm]{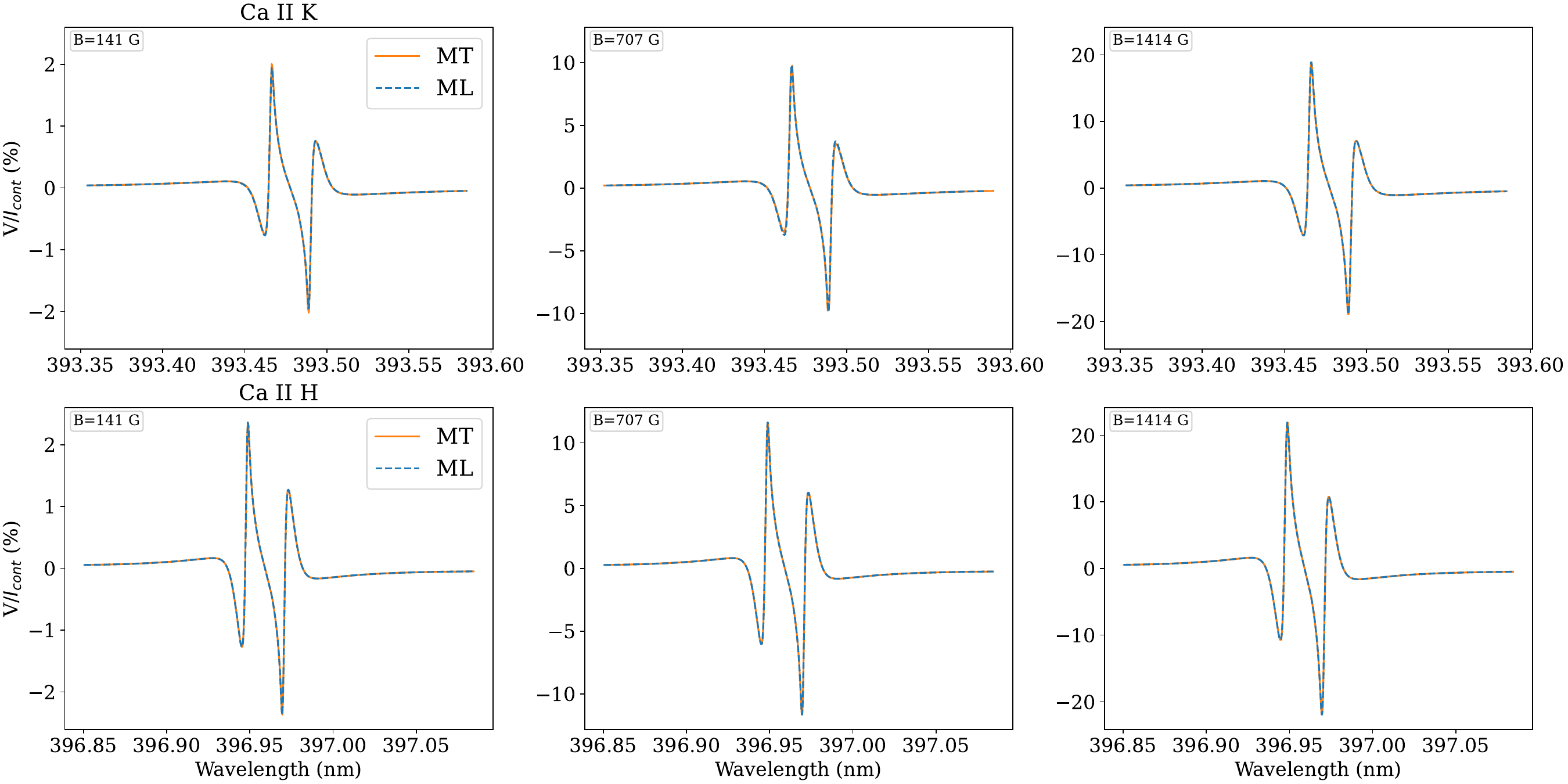}
 \caption{Fractional circular polarisation of the \ion{Ca}{II} H and K lines under the multilevel and multiterm formalisms for different values of $B$ when observed at disc centre. Again, as in Fig.~\ref{Figure:2}, solid curves represent the results of the MT formalism, while the results of the MZ formalism are shown in dashed curves. The panels on the top show the \ion{Ca}{II} K line profiles, and the panels on the bottom show the \ion{Ca}{II} H line profiles.}
 \label{Figure:4}
\end{figure*}

We synthesised the Stokes $V$ profiles for different values of $B_{\mathrm{LoS}}$ for $\mu=1$ using MZ and MT formalisms. The results are presented in Fig.~\ref{Figure:4}. The outputs from both formalisms are virtually identical. As the circular polarisation in the \ion{Ca}{II} H and K lines is predominantly dominated by the Zeeman-induced polarisation, the results for other heliocentric angles bear great resemblance to the result presented for $\mu=1$, so they are not shown here.

The circular polarisation signals of the \ion{Ca}{II} H and K lines are orders of magnitude stronger than the corresponding linear polarisation signals (upper panel of Fig.~\ref{Figure:3}). This large amplitude underscores the potential to use the Stokes $V$ signal for chromospheric magnetic field diagnostics. 

However, the spectral profiles shown in Fig.~\ref{Figure:3} and Fig.~\ref{Figure:4} were computed at a much higher spectral resolution than is typically achievable in observations with instruments such as CHROMIS. Additionally, these profiles do not incorporate the effects of noise, which would further complicate the analysis and reduce the usability of the weaker linear polarisation signals. We investigate these effects in the following sections.

\begin{figure*}
 \centering
 \includegraphics[width=18cm]{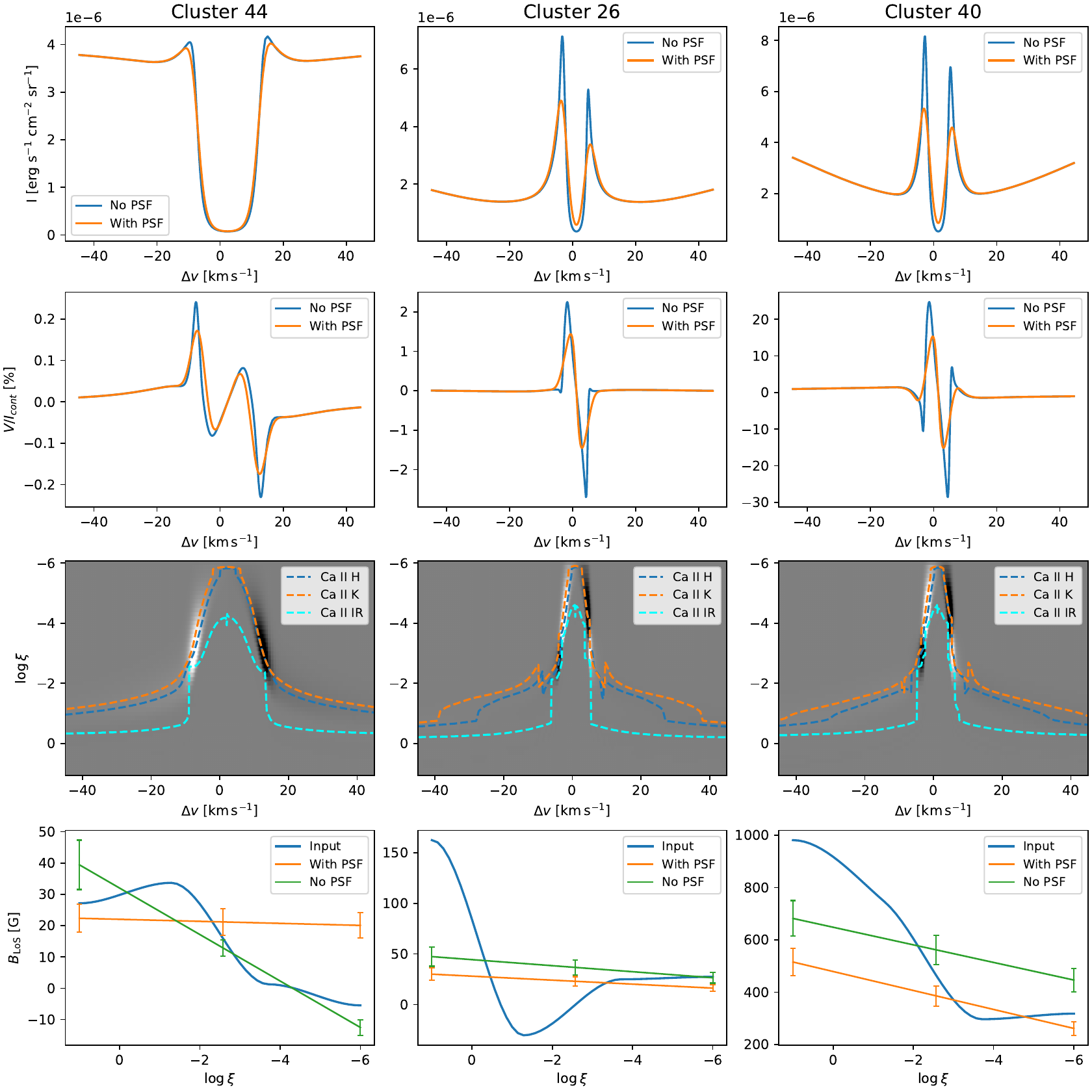} 
 \caption{Line profiles, response functions, and magnetic fields in three representative atmosphere models. 
 Top row: Intensity of the \CaIIH\ line before (blue) and after (orange) being convolved with a spectral PSF of 130 m\AA. 
 Second row: Circular polarisation of the \CaIIH\ line before (blue) and after (orange) being convolved with a spectral PSF of 130 m\AA.
 Third row:  Image of the response functions of Stokes $V$ in the \ion{Ca}{II} H line to variations of $B_{\text{LoS}}$, as function of wavelength expressed as Doppler shift and column mass $\xi$. The three dashed curves correspond to the location on the $\log{\xi}$ vs $\Delta \lambda$ space of the maxima of the response functions for the \ion{Ca}{II} H (blue), K (orange), and 854.2 nm (cyan) lines.
 Bottom row:  Stratification of $B_{\text{LoS}}$ obtained from inverting the synthesized \ion{Ca}{II} H and K spectra compared to the input atmospheric model. Vertical lines on the bottom row represent estimations of the uncertainty of the $B_{\mathrm{LoS}}$ at chosen depth points.}

 \label{Figure:6}%
\end{figure*}

\subsection{Synthetic profiles}

We now turn our attention from the FAL-C atmosphere to the 3D model atmosphere that we constructed from the \CaIIIR\ observations (see Sec.~\ref{subsec:inversion_strategy}).

We used the inverted atmospheres of the 45 representative pixels of each $k$-mean cluster to compute synthetic spectra using STiC of the \ion{Ca}{II} H and K lines. This step allowed us to evaluate the polarisation signals that are expected from different solar structures.

As in the semi-empirical atmosphere discussed in Sect.\ref{sect:falc}, the fractional linear polarisation ($Q/I_{cont}$ and $U/I_{cont}$) in all cases was well below 1\%. Therefore, we focus on the circular polarisation ($V$), whose results are shown for three representative clusters in the second row of Fig.~\ref{Figure:6}, and for all clusters in Fig.~\ref{Figure:A1}.

The synthetic Stokes $V$ profiles display a wide range of behaviours, reflecting the diversity of solar structures sampled by the observations. The maximum circular polarisation fraction spans from about 0.2\% in clusters located in regions of weak line-of-sight magnetic field to values as large as 20\% in magnetic field concentrations. Most clusters lie between these extremes. As expected, pixels with lower polarisation fractions correspond to areas far from photospheric magnetic concentrations, often in fibrillar regions where the magnetic field is largely horizontal. Conversely, the largest polarisation fractions are found in clusters associated with strong magnetic field concentrations, where both the field strength and $B_{\mathrm{LoS}}$ are high.

\subsubsection{Response functions}\label{sec:rf}

To investigate where in the atmosphere the \ion{Ca}{II} H and K lines are sensitive to magnetic fields, we computed numerical response functions of Stokes $V$ to $B_{\text{LoS}}$ with STiC for the 45 representative pixels. The third row of Fig.~\ref{Figure:6} shows the results for three clusters described earlier, while the response functions for all 45 clusters are provided in Fig.~\ref{Figure:A2}. The response functions of the \ion{Ca}{II} K line are nearly identical to those of the H line and are omitted here for clarity.

Most response functions are not antisymmetric with respect to the line core. In fact, at the same wavelength one can often find both positive and negative values at different depths. This behaviour is consistent with the presence of vertical gradients in the source function \citep[][]{2006ASPC..354..313U}. An inspection of the source function as a function of height confirmed that such gradients are indeed present in all the relevant profiles.

In the wings of the lines, the response is mainly to magnetic fields low in the atmosphere (around $\log{\xi} = -1$). Closer to the line cores, the response is much stronger, and Stokes $V$ becomes sensitive to magnetic fields in the upper chromosphere, with a maximum of the response function above $\log{\xi} = -4$. For comparison with the \ion{Ca}{II} 854.2 nm line, the dashed curves in the third row of Fig.~\ref{Figure:6} show the height of the maximum of the absolute value of the response function as a function of wavelength for all three lines. The results clearly indicate that the \ion{Ca}{II} H and K lines (blue and orange curves) probe magnetic fields at greater heights than their infrared counterpart (cyan curve).

Due to its larger opacity, the \CaIIK\ line is sensitive to slightly higher atmospheric layers than the \CaIIH\ line, although the two largely overlap. Assuming hydrostatic equilibrium, we find that at line centre the difference in height of the response-peak locations between the two lines varies between about 90 and 130 km. Consequently, spectropolarimetric observations of both lines can provide not only redundancy but also complementary information, helping to better constrain magnetic field values in the upper chromosphere, provided that the observations achieve sufficient S/N. Comparatively, the difference in height of the response-peak locations between the \CaIIK\ and \ion{Ca}{II} 854.2 nm line is around 400 km.

\subsubsection{\CaIIHK\ inversions} \label{subsubsec:inv}
Building on the results of the response functions, we next tested whether $B_{\mathrm{LoS}}$ can be constrained through inversions of the \ion{Ca}{II} H and K spectropolarimetric profiles. For this purpose, we used STiC to invert the synthetic spectra of the three representative pixels shown in Fig.~\ref{Figure:6}.

Inversions were performed in two cycles: first to constrain the temperature, line-of-sight velocity, and microturbulent velocity, and then a second cycle with one, two, or three nodes in $B_{\mathrm{LoS}}$. The inversion with the lowest value of $\chi$ was chosen as the best fit. To evaluate the impact of instrumental spectral PSF, we performed inversions both with and without convolving the synthetic spectra with the CHROMIS PSF, which has a full width at half-maximum of 0.13~$\AA$. The bottom row of Fig.~\ref{Figure:6} shows, for each pixel, the stratification recovered in $B_{\mathrm{LoS}}$ compared to the original atmosphere. For those pixels, the cycle with two nodes in $B_{\mathrm{LoS}}$ yielded the lowest value of $\chi$. We used the numerical response functions to estimate the uncertainty of the obtained $B_{\mathrm{LoS}}$ at three depth points using the method described in \citet[][]{2003isp..book.....D}.

The results vary depending on the strength of the polarisation signal. For the pixel with weak circular polarisation (left column), inversion without PSF recovers the overall behaviour of $B_{\mathrm{LoS}}$, but the retrieval degrades when PSF is included. For the intermediate case (middle column), where the chromospheric magnetic field is stronger, both inversions reproduce the magnetic field at higher layers reasonably well, although the lower-layer field is not constrained, consistent with the limited sensitivity of the H and K lines to deep layers. Finally, for the pixel with the strongest chromospheric magnetic field and polarisation signal (right column), the inversion with PSF convolution actually outperforms the one without, yielding a better match to the input stratification.

\begin{figure*}
 \centering
 \includegraphics[width=19.5cm]{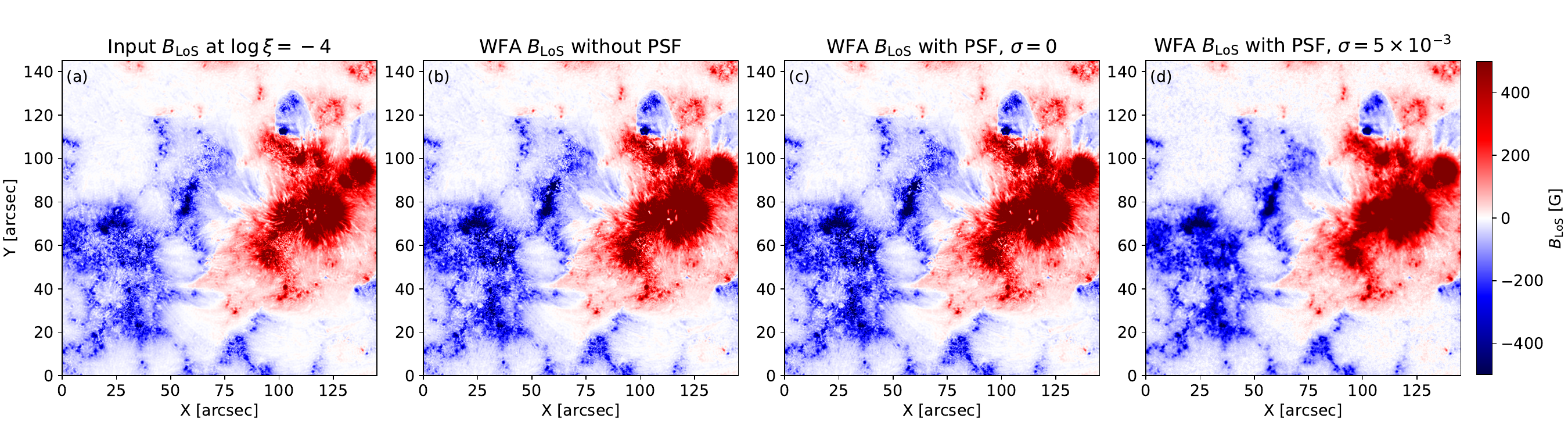}
 \includegraphics[width=19.5cm]{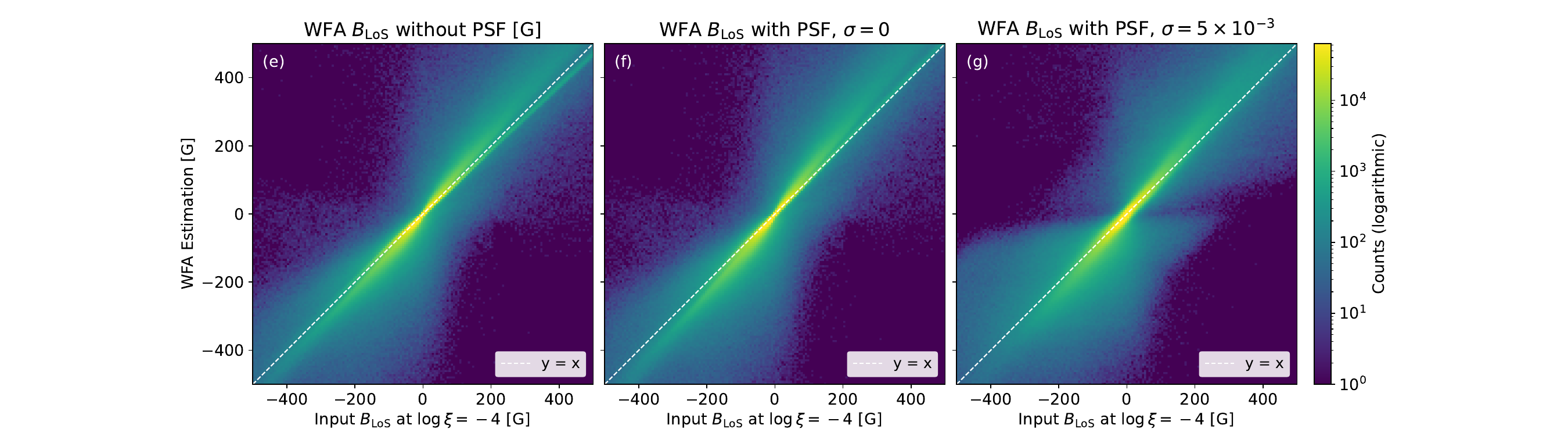}
 \caption{Top row: Line-of-sight magnetic field ($B_{\mathrm{LoS}}$) at $\log{\xi}=-4$ from the input model (a), compared with the weak-field estimates obtained from the original synthetic spectra (b), and from the same spectra after convolution with the CHROMIS spectral PSF with (c), and without noise (d). Bottom row: Logarithmic density plots comparing the input $B_{\mathrm{LoS}}$ with the weak-field results, for the case without the convolution with the PSF (e), and with the convolution with the PSF for the noiseless case (f) and the case with added noise (g). Dashed lines in the bottom row mark the identity line.}
 \label{Figure:8}%
\end{figure*}

\subsubsection{Weak field approximation} \label{sect:wfa}

We showed in Sec.~\ref{subsubsec:inv} that full non-LTE inversions can retrieve a good estimate of the magnetic field in the chromosphere. However, such inversions take  $\sim$1~h of CPU time per pixel, and it would be desirable to obtain an estimate of the field that is quicker to compute. This is particularly important for time series of CHROMIS observations, which contain  $\sim 5 \times 10^{6}$ spatial pixels per line scan.

Therefore, we applied the WFA to the synthetic \CaIIH\ profiles. Based on the response function analysis in Sect.~\ref{sec:rf}, we restricted the calculation to a 0.04 nm window around the \ion{Ca}{II} H line core, where sensitivity to upper-chromospheric magnetic fields is strongest. Weak-field calculations were performed with and without convolving the profiles with the CHROMIS spectral PSF (FWHM = 0.13 Å) using a typical spectral sampling of 0.065~nm. To simulate the conditions of real observations, we also repeated the calculations adding a noise level of $5 \times 10^{-3}$. We used the spatially-regularised WFA method of \citet[][]{2024A&A...685A..85D} to perform the magnetic field inference.

The results are presented in Fig.~\ref{Figure:8}. We used the structural similarity index (SSIM, \citealp{4775883}), which quantifies the similarity between two maps based on their intensity patterns and spatial structure through local mean, variance, and cross-covariance. Our analysis shows that the weak-field estimate of $B_{\mathrm{LoS}}$ matches the input atmosphere at $\log{\xi} \approx -4$ best, consistent with the sensitivity derived from the response functions. The corresponding $B_{\mathrm{LoS}}$ map is shown in panel (a) of Fig.~\ref{Figure:8}.
Comparison of panels (b) and (c) indicates that convolution with the spectral PSF does not introduce significant differences. This is further supported by the logarithmic density plots in panels (e) and (f), which show nearly identical scatter relative to the input atmosphere. However, adding noise, as shown in panels (d) and (g), degrades the quality of the inferred $B_{\mathrm{LoS}}$. Specifically, the density plot in panel (g) exhibits a substantial increase in scatter compared to the noiseless case.
A summary of the density plots is provided in Table~\ref{table:Table2}, which lists various percentile values. For $B_{\mathrm{LoS}}$ inference without PSF convolution and without added noise, the 50\% percentile value is 6~G. This indicates that for 50\% of the pixels, the difference between the inferred $B_{\mathrm{LoS}}$ using the weak-field approximation and the input $B_{\mathrm{LoS}}$ at $\log{\xi} = -4$ is less than 6~G. Table~\ref{table:Table2} confirms that PSF convolution has minimal impact, with the 50\% percentile unchanged and the 70\% percentile increasing by less than 5~G. In contrast, added noise significantly affects the results: the 50\% percentile is nearly three times larger than the noiseless case, and the 70\% percentile is almost twice as large.

 \begin{table}
    \renewcommand*{\arraystretch}{1.5}
    \centering
    \caption{Percentile values (in G) for the density plots of Fig.~\ref{Figure:8}.}
    \label{table:Table2}
    \begin{tabular}{c|c|c|c|}
        \cline{2-4}
        & \multicolumn{3}{c|}{Percentile} \\ \cline{2-4}
        & 50\% & 70\% & 90\% \\            
        \cline{1-4}
        \multicolumn{1}{|c|}{Without PSF, $\sigma =0.0$} & 6 & 15 & 64 \\
       \multicolumn{1}{|c|}{ With PSF, $\sigma =0.0$} & 6 & 20 & 75 \\
        \multicolumn{1}{|c|}{With PSF, $\sigma = 5\times 10^{-3}$} & 17 & 30 & 90 \\
        \cline{1-4}
    \end{tabular}
\end{table}

\section{Conclusions} \label{conclusions}

In this work, we explored the diagnostic potential of the \ion{Ca}{II} H and K lines to probe magnetic fields in the solar chromosphere. To this end, we combined forward modelling and inversions based on both semi-empirical atmospheric models and atmospheres retrieved from high-resolution observations.

A first step was to assess the validity of the MZ approximation compared with the more complete MT formalism. We confirmed that the MZ approximation is inadequate for reproducing the formation of linear polarisation in these lines, consistent with earlier studies \citep[][]{1980A&A....84...68S,2011ApJ...743....3B,2025arXiv251019719J}. However, we also showed that circular polarisation can be reliably modelled  under the MZ formalism, meaning that widely used inversion codes such as STiC \citep{2019A&A...623A..74D} can be safely applied to this observable.

From both semi-empirical and inverted model atmospheres, we synthesised spectropolarimetric profiles and examined their polarisation properties. The linear polarisation amplitudes are weak (below 1.7\% in our tests), making them hard to detect with an instrument like CHROMIS. The circular polarisation, in contrast, shows a wide range of amplitudes, from fractions of a percent in quiet-Sun conditions to more than 10\% in strong magnetic concentrations, even if the observations have a spectral resolution of 30,000.

The response functions to $B_{\text{LoS}}$ showed that the wings of the \ion{Ca}{II} H and K lines are sensitive to magnetic fields in the lower chromosphere (around $\log{\xi}=-1$), while their cores respond to magnetic fields higher in the chromosphere, peaks above $\log{\xi}=-4$. Compared to the commonly used \ion{Ca}{II} 854.2 nm line, the \ion{Ca}{II} H and K lines probe higher atmospheric layers. 

Although the peak response depths of the \ion{Ca}{II}~H and K lines largely overlap, they are not identical: assuming hydrostatic equilibrium, the location of the peak response at line core differs by approximately 90 to 130~km. This offset reflects intrinsic differences between the two transitions: the K line, having the larger oscillator strength, and thus opacity, tends to form slightly higher in the atmosphere, whereas the H line has a larger effective Landé factor and a marginally larger wavelength. Consequently, the Zeeman splitting, which scales as $\Delta\lambda_{\mathrm{B}} \propto g_{\mathrm{eff}} \lambda^{2} B$, is somewhat larger for the H line, providing a modest enhancement in magnetic sensitivity per unit of field strength. In practice, the K line exhibits stronger line depression but lower photon flux, while the H line offers slightly larger S/N under comparable observing conditions (see Fig.~\ref{Figure:A1}).

We further tested the feasibility of constraining $B_{\text{LoS}}$ through inversions of synthetic \ion{Ca}{II} H and K line profiles. For representative pixels, the inversions were able to recover the magnetic field in the upper chromosphere, even when the profiles were degraded by the CHROMIS spectral PSF. To extend the analysis to the full field of view, we inverted the \ion{Ca}{II} 854.2 nm observations with STiC. The resulting atmospheric models were used to synthesise \ion{Ca}{II} H and K line profiles, to which we applied the weak-field approximation. Using the structure similarity index, we found that the weak-field estimate $B_{\text{LoS}}$ best matches the input atmosphere around $\log{\xi}=-4$, consistent with the response function analysis. Even at spectral resolution, the number of spectral samples and noise level that we expect to obtain with CHROMIS, $B_{\text{LoS}}$ are retrieved well.

Together, these results establish that the \ion{Ca}{II} H and K lines offer sensitivity to line-of-sight magnetic fields in the upper chromosphere, at heights inaccessible with the popular chromospheric diagnostics obtained from the \ion{Ca}{II} 854.2 nm line
\citep[e.g.][]{2019ApJ...874..126K,2020A&A...642A.210M},
while also being observable from the ground, in contrast to the \ion{Mg}{II} h and k lines
\citep{2021SciA....7.8406I}. 

The large amplitude of the circular polarisation signal indicates that the CHROMIS instrument will be able to use the \CaIIHK\ lines for inference of the line-of-sight magnetic field. However, the linear polarisation signals are too weak to reliably obtain estimates of the magnetic field in the plane of sky .

\begin{acknowledgements}
The authors of this paper gratefully acknowledge funding by the European Union through the European Research Council (ERC) under the Horizon Europe program (MAGHEAT, grant agreement 101088184), and the Swedish Research Council (registration number 2022-03535 and 2021-05613), and Swedish National Space Agency (2021-00116).
The Swedish 1-m Solar Telescope is operated on the island of La Palma by the Institute for Solar Physics of Stockholm University in the Spanish Observatorio del Roque de los Muchachos of the Instituto de Astrof\'\i sica de Canarias. The Institute for Solar Physics is supported by a grant for research infrastructures of national importance from the Swedish Research Council (registration number 2021-00169). The computations were enabled by resources provided by the National Academic Infrastructure for Supercomputing in Sweden (NAISS), partially funded by the Swedish Research Council through grant agreement no. 2022-06725. The authors acknowledge the National Academic Infrastructure for Supercomputing in Sweden (NAISS), partially funded by the Swedish Research Council through grant agreement no. 2022-06725, for awarding this project access to the LUMI supercomputer, owned by the EuroHPC Joint Undertaking and hosted by CSC (Finland) and the LUMI consortium.
\end{acknowledgements}

\bibliographystyle{aa}
\bibliography{citations}

\begin{thebibliography}{50}
\expandafter\ifx\csname natexlab\endcsname\relax\def\natexlab#1{#1}\fi

\bibitem[{{Avrett}(1985)}]{1985cdm..proc...67A}
{Avrett}, E.~H. 1985, in Chromospheric Diagnostics and Modelling, ed. B.~W. {Lites}, 67--127

\bibitem[{{Belluzzi} \& {Trujillo Bueno}(2011)}]{2011ApJ...743....3B}
{Belluzzi}, L. \& {Trujillo Bueno}, J. 2011, \apj, 743, 3

\bibitem[{{Bj{\o}rgen} {et~al.}(2018){Bj{\o}rgen}, {Sukhorukov}, {Leenaarts}, {Carlsson}, {de la Cruz Rodr{\'\i}guez}, {Scharmer}, \& {Hansteen}}]{2018A&A...611A..62B}
{Bj{\o}rgen}, J.~P., {Sukhorukov}, A.~V., {Leenaarts}, J., {et~al.} 2018, \aap, 611, A62

\bibitem[{{Casini} {et~al.}(2014){Casini}, {Landi Degl'Innocenti}, {Manso Sainz}, {Landi Degl'Innocenti}, \& {Landolfi}}]{2014ApJ...791...94C}
{Casini}, R., {Landi Degl'Innocenti}, M., {Manso Sainz}, R., {Landi Degl'Innocenti}, E., \& {Landolfi}, M. 2014, \apj, 791, 94

\bibitem[{{de la Cruz Rodr{\'\i}guez}(2019)}]{2019A&A...631A.153D}
{de la Cruz Rodr{\'\i}guez}, J. 2019, \aap, 631, A153

\bibitem[{{de la Cruz Rodr{\'\i}guez} \& {Leenaarts}(2024)}]{2024A&A...685A..85D}
{de la Cruz Rodr{\'\i}guez}, J. \& {Leenaarts}, J. 2024, \aap, 685, A85

\bibitem[{{de la Cruz Rodr{\'\i}guez} {et~al.}(2016){de la Cruz Rodr{\'\i}guez}, {Leenaarts}, \& {Asensio Ramos}}]{2016ApJ...830L..30D}
{de la Cruz Rodr{\'\i}guez}, J., {Leenaarts}, J., \& {Asensio Ramos}, A. 2016, \apjl, 830, L30

\bibitem[{{de la Cruz Rodr{\'\i}guez} {et~al.}(2019){de la Cruz Rodr{\'\i}guez}, {Leenaarts}, {Danilovic}, \& {Uitenbroek}}]{2019A&A...623A..74D}
{de la Cruz Rodr{\'\i}guez}, J., {Leenaarts}, J., {Danilovic}, S., \& {Uitenbroek}, H. 2019, \aap, 623, A74

\bibitem[{{de la Cruz Rodr{\'\i}guez} {et~al.}(2015){de la Cruz Rodr{\'\i}guez}, {L{\"o}fdahl}, {S{\"u}tterlin}, {Hillberg}, \& {Rouppe van der Voort}}]{2015A&A...573A..40D}
{de la Cruz Rodr{\'\i}guez}, J., {L{\"o}fdahl}, M.~G., {S{\"u}tterlin}, P., {Hillberg}, T., \& {Rouppe van der Voort}, L. 2015, \aap, 573, A40

\bibitem[{{de la Cruz Rodr{\'\i}guez} \& {Piskunov}(2013)}]{2013ApJ...764...33D}
{de la Cruz Rodr{\'\i}guez}, J. \& {Piskunov}, N. 2013, \apj, 764, 33

\bibitem[{{de Wijn} {et~al.}(2022){de Wijn}, {Casini}, {Carlile}, {Lecinski}, {Sewell}, {Zmarzly}, {Eigenbrot}, {Beck}, {W{\"o}ger}, \& {Kn{\"o}lker}}]{2022SoPh..297...22D}
{de Wijn}, A.~G., {Casini}, R., {Carlile}, A., {et~al.} 2022, \solphys, 297, 22

\bibitem[{{del Pino Alem{\'a}n} {et~al.}(2016){del Pino Alem{\'a}n}, {Casini}, \& {Manso Sainz}}]{2016ApJ...830L..24D}
{del Pino Alem{\'a}n}, T., {Casini}, R., \& {Manso Sainz}, R. 2016, \apjl, 830, L24

\bibitem[{{del Toro Iniesta}(2003)}]{2003isp..book.....D}
{del Toro Iniesta}, J.~C. 2003, {Introduction to Spectropolarimetry}

\bibitem[{{Feller} {et~al.}(2020){Feller}, {Gandorfer}, {Iglesias}, {Lagg}, {Riethm{\"u}ller}, {Solanki}, {Katsukawa}, \& {Kubo}}]{2020SPIE11447E..AKF}
{Feller}, A., {Gandorfer}, A., {Iglesias}, F.~A., {et~al.} 2020, in Society of Photo-Optical Instrumentation Engineers (SPIE) Conference Series, Vol. 11447, Ground-based and Airborne Instrumentation for Astronomy VIII, ed. C.~J. {Evans}, J.~J. {Bryant}, \& K.~{Motohara}, 11447AK

\bibitem[{{Fontenla} {et~al.}(1993){Fontenla}, {Avrett}, \& {Loeser}}]{1993ApJ...406..319F}
{Fontenla}, J.~M., {Avrett}, E.~H., \& {Loeser}, R. 1993, \apj, 406, 319

\bibitem[{{Harvey}(2012)}]{2012SoPh..280...69H}
{Harvey}, J.~W. 2012, \solphys, 280, 69

\bibitem[{{Ishikawa} {et~al.}(2021){Ishikawa}, {Trujillo Bueno}, {del Pino Alem{\'a}n}, {Okamoto}, {McKenzie}, {Auch{\`e}re}, {Kano}, {Song}, {Yoshida}, {Rachmeler}, {Kobayashi}, {Hara}, {Kubo}, {Narukage}, {Sakao}, {Shimizu}, {Suematsu}, {Bethge}, {De Pontieu}, {Sainz Dalda}, {Vigil}, {Winebarger}, {Alsina Ballester}, {Belluzzi}, {{\v{S}}t{\v{e}}p{\'a}n}, {Ramos}, {Carlsson}, \& {Leenaarts}}]{2021SciA....7.8406I}
{Ishikawa}, R., {Trujillo Bueno}, J., {del Pino Alem{\'a}n}, T., {et~al.} 2021, Science Advances, 7, eabe8406

\bibitem[{{Juanikorena Berasategi} {et~al.}(2025){Juanikorena Berasategi}, {Alsina Ballester}, {del Pino Alem{\'a}n}, \& {Trujillo Bueno}}]{2025A&A...704A.173J}
{Juanikorena Berasategi}, I., {Alsina Ballester}, E., {del Pino Alem{\'a}n}, T., \& {Trujillo Bueno}, J. 2025, \aap, 704, A173

\bibitem[{{Kleint}(2012)}]{2012ApJ...748..138K}
{Kleint}, L. 2012, \apj, 748, 138

\bibitem[{{Korpi-Lagg} {et~al.}(2025){Korpi-Lagg}, {Gandorfer}, {Solanki}, {del Toro Iniesta}, {Katsukawa}, {Bernasconi}, {Berkefeld}, {Feller}, {Riethm{\"u}ller}, {{\'A}lvarez-Herrero}, {Kubo}, {Mart{\'\i}nez Pillet}, {Smitha}, {Orozco Su{\'a}rez}, {Grauf}, {Carpenter}, {Bell}, {{\'A}lvarez-Alonso}, {{\'A}lvarez Garc{\'\i}a}, {Aparicio del Moral}, {Ati{\'e}nzar}, {Ayoub}, {Bail{\'e}n}, {Bail{\'o}n Mart{\'\i}nez}, {Balaguer Jim{\'e}nez}, {Barthol}, {Bayon Laguna}, {Bellot Rubio}, {Bergmann}, {Blanco Rodr{\'\i}guez}, {Bochmann}, {Borrero}, {Campos-Jara}, {Castellanos Dur{\'a}n}, {Cebollero}, {Conde Rodr{\'\i}guez}, {Deutsch}, {Eaton}, {Fern{\'a}ndez-Medina}, {Fernandez-Rico}, {Ferreres}, {Garc{\'\i}a}, {Garc{\'\i}a Alarcia}, {Garc{\'\i}a Parejo}, {Garranzo-Garc{\'\i}a}, {Gasent Blesa}, {Gerber}, {Germerott}, {Gilabert Palmer}, {Gizon}, {G{\'o}mez S{\'a}nchez-Tirado}, {Gonz{\'a}lez-B{\'a}rcena}, {Gonzalo Melchor}, {Goodyear}, {Hara}, {Harnes}, {Heerlein}, {Heidecke}, {Heinrichs}, {Hern{\'a}ndez Exp{\'o}sito},
  {Hirzberger}, {Hoelken}, {Hyun}, {Iglesias}, {Ishikawa}, {Jeon}, {Kawabata}, {Kolleck}, {Laguna}, {Lomas}, {L{\'o}pez Jim{\'e}nez}, {Manzano}, {Matsumoto}, {Mayo Turrado}, {Meierdierks}, {Meining}, {Monecke}, {Morales-Fern{\'a}ndez}, {Moreno Mantas}, {Moreno Vacas}, {M{\"u}ller}, {M{\"u}ller}, {Naito}, {Nakai}, {N{\'u}{\~n}ez Peral}, {Oba}, {Palo}, {P{\'e}rez-Grande}, {Piqueras Carre{\~n}o}, {Preis}, {Przybylski}, {Quintero Noda}, {Ramanath}, {Ramos M{\'a}s}, {Raouafi}, {Rivas-Mart{\'\i}nez}, {Rodr{\'\i}guez Mart{\'\i}nez}, {Rodr{\'\i}guez Valido}, {Ruiz Cobo}, {S{\'a}nchez Rodr{\'\i}guez}, {Sanchez Toledo}, {S{\'a}nchez G{\'o}mez}, {Sanchis Kilders}, {Sant}, {Santamarina Guerrero}, {Schulze}, {Shimizu}, {Silva-L{\'o}pez}, {Singh}, {Siu-Tapia}, {Sonner}, {Staub}, {Strecker}, {Tobaruela}, {Torralbo}, {Tritschler}, {Tsuzuki}, {Uraguchi}, {Volkmer}, {Vourlidas}, {Vukadinovi{\'c}}, {Werner}, \& {Zerr}}]{2025SoPh..300...75K}
{Korpi-Lagg}, A., {Gandorfer}, A., {Solanki}, S.~K., {et~al.} 2025, \solphys, 300, 75

\bibitem[{{Kriginsky} \& {Oliver}(2025)}]{2025ApJ...981..121K}
{Kriginsky}, M. \& {Oliver}, R. 2025, \apj, 981, 121

\bibitem[{{Kriginsky} {et~al.}(2021){Kriginsky}, {Oliver}, {Antolin}, {Kuridze}, \& {Freij}}]{2021A&A...650A..71K}
{Kriginsky}, M., {Oliver}, R., {Antolin}, P., {Kuridze}, D., \& {Freij}, N. 2021, \aap, 650, A71

\bibitem[{{Kriginsky} {et~al.}(2020){Kriginsky}, {Oliver}, {Freij}, {Kuridze}, {Asensio Ramos}, \& {Antolin}}]{2020A&A...642A..61K}
{Kriginsky}, M., {Oliver}, R., {Freij}, N., {et~al.} 2020, \aap, 642, A61

\bibitem[{{Kuridze} {et~al.}(2019){Kuridze}, {Mathioudakis}, {Morgan}, {Oliver}, {Kleint}, {Zaqarashvili}, {Reid}, {Koza}, {L{\"o}fdahl}, {Hillberg}, {Kukhianidze}, \& {Hanslmeier}}]{2019ApJ...874..126K}
{Kuridze}, D., {Mathioudakis}, M., {Morgan}, H., {et~al.} 2019, \apj, 874, 126

\bibitem[{{Landi Degl'Innocenti} \& {Landolfi}(2004)}]{2004ASSL..307.....L}
{Landi Degl'Innocenti}, E. \& {Landolfi}, M. 2004, {Polarization in Spectral Lines}, Vol. 307

\bibitem[{{Leenaarts} {et~al.}(2018){Leenaarts}, {de la Cruz Rodr{\'\i}guez}, {Danilovic}, {Scharmer}, \& {Carlsson}}]{2018A&A...612A..28L}
{Leenaarts}, J., {de la Cruz Rodr{\'\i}guez}, J., {Danilovic}, S., {Scharmer}, G., \& {Carlsson}, M. 2018, \aap, 612, A28

\bibitem[{{Leenaarts} {et~al.}(2012){Leenaarts}, {Pereira}, \& {Uitenbroek}}]{2012A&A...543A.109L}
{Leenaarts}, J., {Pereira}, T., \& {Uitenbroek}, H. 2012, \aap, 543, A109

\bibitem[{{Li} {et~al.}(2024){Li}, {del Pino Alem{\'a}n}, \& {Trujillo Bueno}}]{2024ApJ...975..110L}
{Li}, H., {del Pino Alem{\'a}n}, T., \& {Trujillo Bueno}, J. 2024, \apj, 975, 110

\bibitem[{{Li} {et~al.}(2022){Li}, {del Pino Alem{\'a}n}, {Trujillo Bueno}, \& {Casini}}]{2022ApJ...933..145L}
{Li}, H., {del Pino Alem{\'a}n}, T., {Trujillo Bueno}, J., \& {Casini}, R. 2022, \apj, 933, 145

\bibitem[{{L{\"o}fdahl}(2002)}]{2002SPIE.4792..146L}
{L{\"o}fdahl}, M.~G. 2002, in Society of Photo-Optical Instrumentation Engineers (SPIE) Conference Series, Vol. 4792, Image Reconstruction from Incomplete Data, ed. P.~J. {Bones}, M.~A. {Fiddy}, \& R.~P. {Millane}, 146--155

\bibitem[{{L{\"o}fdahl} {et~al.}(2021){L{\"o}fdahl}, {Hillberg}, {de la Cruz Rodr{\'\i}guez}, {Vissers}, {Andriienko}, {Scharmer}, {Haugan}, \& {Fredvik}}]{2021A&A...653A..68L}
{L{\"o}fdahl}, M.~G., {Hillberg}, T., {de la Cruz Rodr{\'\i}guez}, J., {et~al.} 2021, \aap, 653, A68

\bibitem[{{Martinez Pillet} {et~al.}(1990){Martinez Pillet}, {Garcia Lopez}, {del Toro Iniesta}, {Rebolo}, {Vazquez}, {Beckman}, \& {Char}}]{1990ApJ...361L..81M}
{Martinez Pillet}, V., {Garcia Lopez}, R.~J., {del Toro Iniesta}, J.~C., {et~al.} 1990, \apjl, 361, L81

\bibitem[{{Morosin} {et~al.}(2020){Morosin}, {de la Cruz Rodr{\'\i}guez}, {Vissers}, \& {Yadav}}]{2020A&A...642A.210M}
{Morosin}, R., {de la Cruz Rodr{\'\i}guez}, J., {Vissers}, G. J.~M., \& {Yadav}, R. 2020, \aap, 642, A210

\bibitem[{{Pietarila} {et~al.}(2007){Pietarila}, {Socas-Navarro}, \& {Bogdan}}]{2007ApJ...663.1386P}
{Pietarila}, A., {Socas-Navarro}, H., \& {Bogdan}, T. 2007, \apj, 663, 1386

\bibitem[{{Piskunov} \& {Valenti}(2017)}]{2017A&A...597A..16P}
{Piskunov}, N. \& {Valenti}, J.~A. 2017, \aap, 597, A16

\bibitem[{{Quintero Noda} {et~al.}(2016){Quintero Noda}, {Shimizu}, {de la Cruz Rodr{\'\i}guez}, {Katsukawa}, {Ichimoto}, {Anan}, \& {Suematsu}}]{2016MNRAS.459.3363Q}
{Quintero Noda}, C., {Shimizu}, T., {de la Cruz Rodr{\'\i}guez}, J., {et~al.} 2016, \mnras, 459, 3363

\bibitem[{{Rimmele} {et~al.}(2020){Rimmele}, {Warner}, {Keil}, {Goode}, {Kn{\"o}lker}, {Kuhn}, {Rosner}, {McMullin}, {Casini}, {Lin}, {W{\"o}ger}, {von der L{\"u}he}, {Tritschler}, {Davey}, {de Wijn}, {Elmore}, {Fehlmann}, {Harrington}, {Jaeggli}, {Rast}, {Schad}, {Schmidt}, {Mathioudakis}, {Mickey}, {Anan}, {Beck}, {Marshall}, {Jeffers}, {Oschmann}, {Beard}, {Berst}, {Cowan}, {Craig}, {Cross}, {Cummings}, {Donnelly}, {de Vanssay}, {Eigenbrot}, {Ferayorni}, {Foster}, {Galapon}, {Gedrites}, {Gonzales}, {Goodrich}, {Gregory}, {Guzman}, {Guzzo}, {Hegwer}, {Hubbard}, {Hubbard}, {Johansson}, {Johnson}, {Liang}, {Liang}, {McQuillen}, {Mayer}, {Newman}, {Onodera}, {Phelps}, {Puentes}, {Richards}, {Rimmele}, {Sekulic}, {Shimko}, {Simison}, {Smith}, {Starman}, {Sueoka}, {Summers}, {Szabo}, {Szabo}, {Wampler}, {Williams}, \& {White}}]{2020SoPh..295..172R}
{Rimmele}, T.~R., {Warner}, M., {Keil}, S.~L., {et~al.} 2020, \solphys, 295, 172

\bibitem[{{Scharmer}(2017)}]{2017psio.confE..85S}
{Scharmer}, G. 2017, in SOLARNET IV: The Physics of the Sun from the Interior to the Outer Atmosphere, 85

\bibitem[{{Scharmer} {et~al.}(2003){Scharmer}, {Bjelksjo}, {Korhonen}, {Lindberg}, \& {Petterson}}]{sst}
{Scharmer}, G.~B., {Bjelksjo}, K., {Korhonen}, T.~K., {Lindberg}, B., \& {Petterson}, B. 2003, 4853

\bibitem[{{Scharmer} {et~al.}(2026){Scharmer}, {de la Cruz Rodr{\'\i}guez}, {Leenaarts}, {Lindberg}, {S{\"u}tterlin}, {Hillberg}, {Pietraszewski}, {de Wijn}, {Foster}, \& {Storey}}]{2026A&A...705A..55S}
{Scharmer}, G.~B., {de la Cruz Rodr{\'\i}guez}, J., {Leenaarts}, J., {et~al.} 2026, \aap, 705, A55

\bibitem[{{Scharmer} {et~al.}(2008){Scharmer}, {Narayan}, {Hillberg}, {de la Cruz Rodriguez}, {L{\"o}fdahl}, {Kiselman}, {S{\"u}tterlin}, {van Noort}, \& {Lagg}}]{2008ApJ...689L..69S}
{Scharmer}, G.~B., {Narayan}, G., {Hillberg}, T., {et~al.} 2008, \apjl, 689, L69

\bibitem[{Scharmer {et~al.}(2008)Scharmer, Narayan, Hillberg, de~la Cruz~Rodriguez, Löfdahl, Kiselman, Sütterlin, van Noort, \& Lagg}]{Scharmer_2008}
Scharmer, G.~B., Narayan, G., Hillberg, T., {et~al.} 2008, The Astrophysical Journal, 689, L69

\bibitem[{{Stenflo}(1980)}]{1980A&A....84...68S}
{Stenflo}, J.~O. 1980, \aap, 84, 68

\bibitem[{{Uitenbroek}(2001)}]{2001ApJ...557..389U}
{Uitenbroek}, H. 2001, \apj, 557, 389

\bibitem[{{Uitenbroek}(2006)}]{2006ASPC..354..313U}
{Uitenbroek}, H. 2006, in Astronomical Society of the Pacific Conference Series, Vol. 354, Solar MHD Theory and Observations: A High Spatial Resolution Perspective, ed. J.~{Leibacher}, R.~F. {Stein}, \& H.~{Uitenbroek}, 313

\bibitem[{Van~Noort {et~al.}(2005)Van~Noort, Der~Voort, \& L{\"o}fdahl}]{VanNoort2005}
Van~Noort, M., Der~Voort, L. R.~V., \& L{\"o}fdahl, M.~G. 2005, Solar Physics, 228, 191

\bibitem[{{Vissers} {et~al.}(2022){Vissers}, {Danilovic}, {Zhu}, {Leenaarts}, {D{\'\i}az Baso}, {da Silva Santos}, {de la Cruz Rodr{\'\i}guez}, \& {Wiegelmann}}]{2022A&A...662A..88V}
{Vissers}, G.~J.~M., {Danilovic}, S., {Zhu}, X., {et~al.} 2022, \aap, 662, A88

\bibitem[{Wang \& Bovik(2009)}]{4775883}
Wang, Z. \& Bovik, A.~C. 2009, IEEE Signal Processing Magazine, 26, 98

\bibitem[{{Wiehr} \& {Stellmacher}(1991)}]{1991A&A...247..379W}
{Wiehr}, E. \& {Stellmacher}, G. 1991, \aap, 247, 379

\bibitem[{{Yadav} {et~al.}(2021){Yadav}, {D{\'\i}az Baso}, {de la Cruz Rodr{\'\i}guez}, {Calvo}, \& {Morosin}}]{2021A&A...649A.106Y}
{Yadav}, R., {D{\'\i}az Baso}, C.~J., {de la Cruz Rodr{\'\i}guez}, J., {Calvo}, F., \& {Morosin}, R. 2021, \aap, 649, A106

\end{thebibliography}

\begin{appendix}
\onecolumn
\section{K-means groups inversions and response functions}

\begin{figure*}[h!]
 \centering

 \includegraphics[width=18cm]{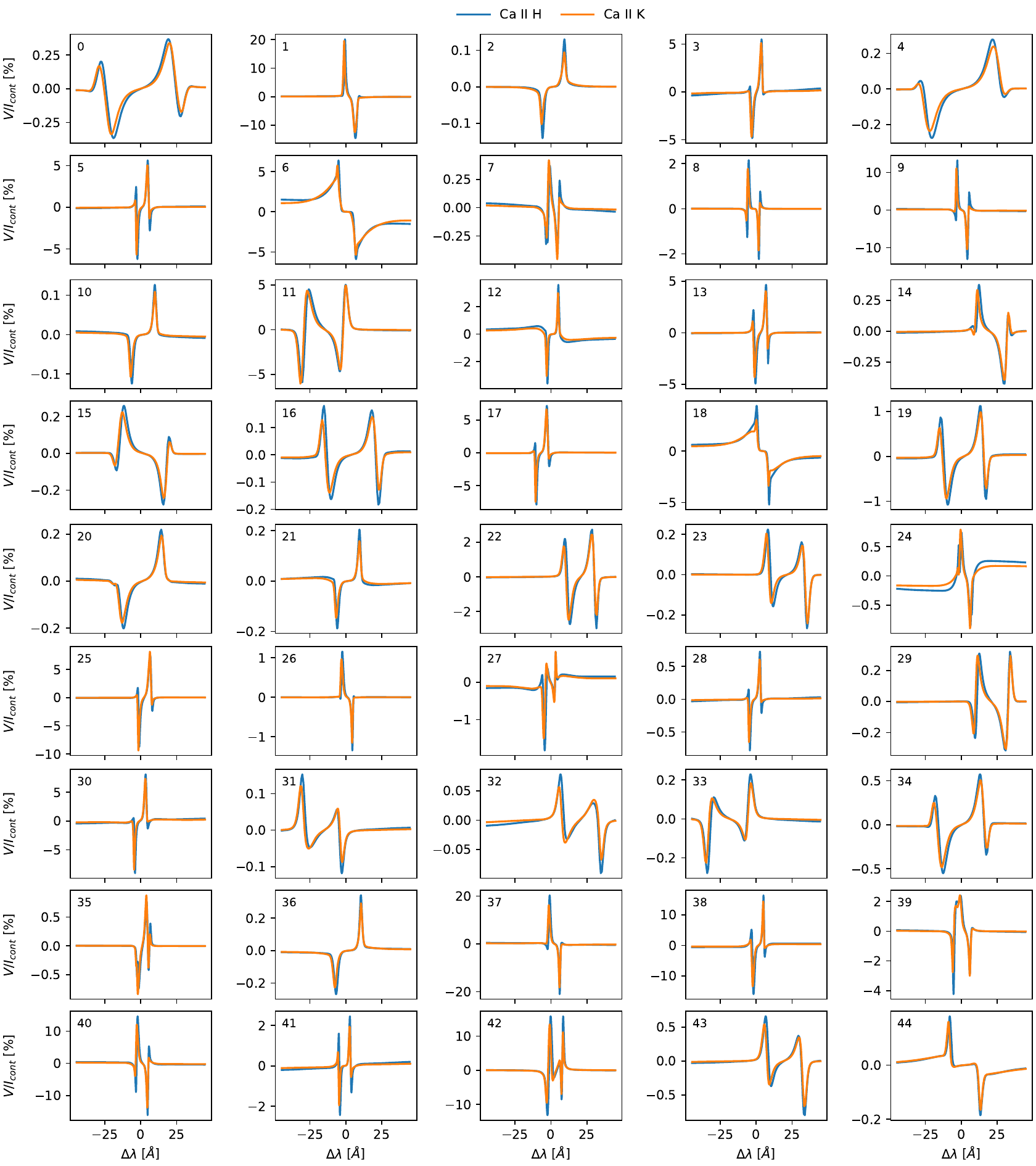}

 \caption{Expanded version of the top row of Fig.~\ref{Figure:6}, including all the identified clusters with their respective \CaIIHK line profiles. The cluster numbers are shown in black inside each panel.}
 \label{Figure:A1}%
\end{figure*}

\begin{figure*}[h!]
 \centering

 \includegraphics[width=18cm]{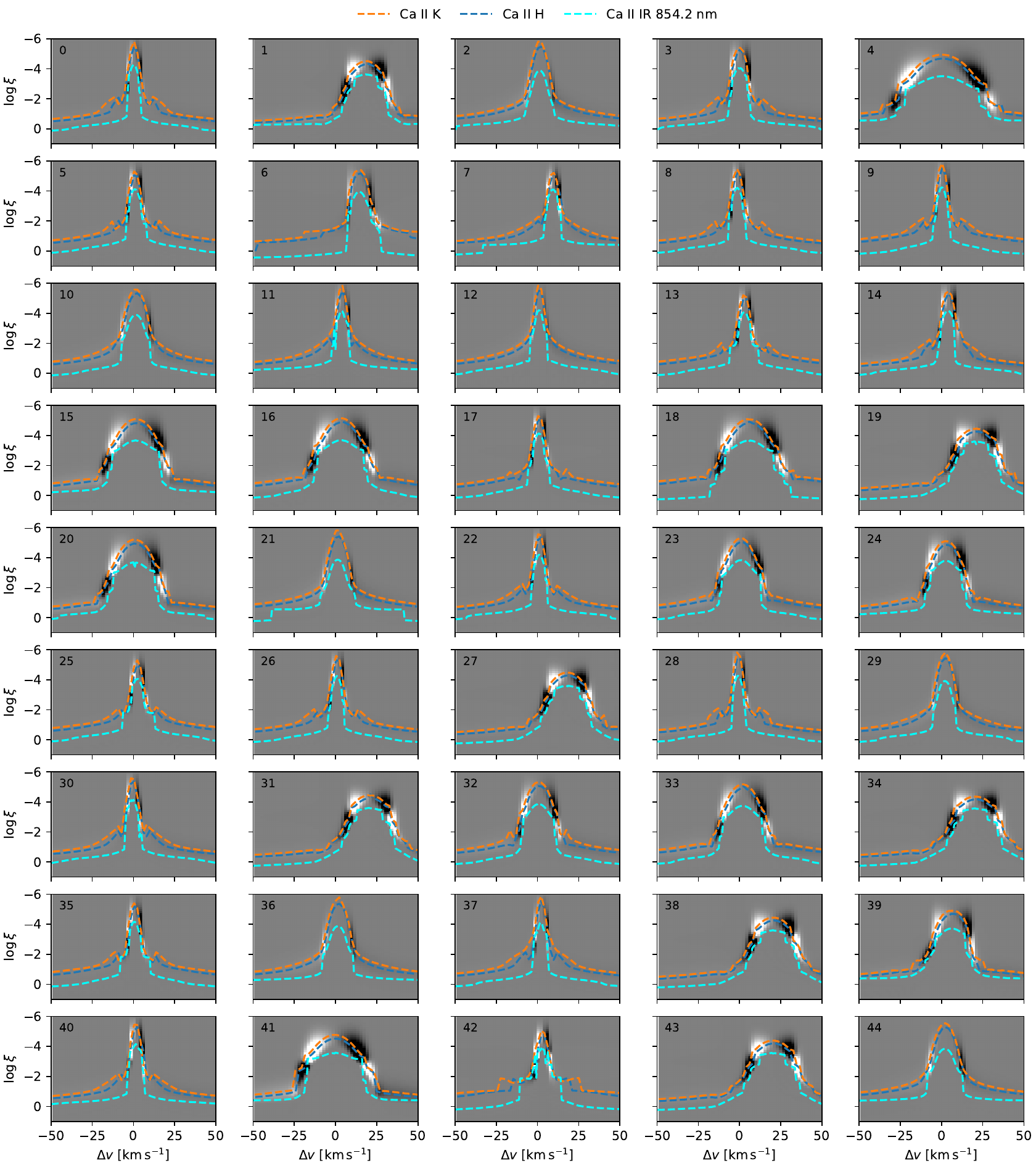}

 \caption{Expanded version of the third row of Fig.~\ref{Figure:6}, including all the identified clusters. The cluster numbers are shown in black inside each panel.}
 \label{Figure:A2}%
\end{figure*}

\end{appendix}

\end{document}